\documentclass[aps,prd,twocolumn,showpacs,preprintnumbers,nofootinbib,amsmath,amssymb]{revtex4}

\usepackage{graphicx}
\usepackage{dcolumn}
\usepackage{bm}
\usepackage{amsmath}
\usepackage[normalem]{ulem}

\newcommand{\beq}{\begin{eqnarray}}
\newcommand{\eeq}{\end{eqnarray}}

\def\spose#1{\hbox to 0pt{#1\hss}}
\def\ltapprox{\mathrel{\spose{\lower 3pt\hbox{$\mathchar"218$}}
 \raise 2.0pt\hbox{$\mathchar"13C$}}}

\begin{document}

\title{Roberge-Weiss endpoint at the physical point of $N_f = 2+1$ QCD}

\author{Claudio Bonati}
\email{claudio.bonati@df.unipi.it}
\affiliation{Dipartimento di Fisica dell'Universit\`a
di Pisa and INFN - Sezione di Pisa,\\ Largo Pontecorvo 3, I-56127 Pisa, Italy}

\author{Massimo D'Elia}
\email{massimo.delia@unipi.it}
\affiliation{Dipartimento di Fisica dell'Universit\`a
di Pisa and INFN - Sezione di Pisa,\\ Largo Pontecorvo 3, I-56127 Pisa, Italy}

\author{Marco Mariti}
\email{mariti@df.unipi.it}
\affiliation{Dipartimento di Fisica dell'Universit\`a
di Pisa and INFN - Sezione di Pisa,\\ Largo Pontecorvo 3, I-56127 Pisa, Italy}

\author{Michele Mesiti}
\email{mesiti@pi.infn.it}
\affiliation{Dipartimento di Fisica dell'Universit\`a
di Pisa and INFN - Sezione di Pisa,\\ Largo Pontecorvo 3, I-56127 Pisa, Italy}

\author{Francesco Negro}
\email{fnegro@pi.infn.it}
\affiliation{Dipartimento di Fisica dell'Universit\`a
di Pisa and INFN - Sezione di Pisa,\\ Largo Pontecorvo 3, I-56127 Pisa, Italy}

\author{Francesco Sanfilippo}
\email{f.sanfilippo@soton.ac.uk}
\affiliation{School of Physics and Astronomy, University of Southampton, SO17 1BJ Southampton, 
United Kindgdom}

\date{\today}

\begin{abstract}
We study the phase diagram of $N_f = 2+1$ QCD in the $T - \mu_B$
plane and investigate the critical point corresponding to the onset of
the Roberge-Weiss transition, which is found for imaginary values of
$\mu_B$. We make use of stout improved staggered fermions and of the
tree level Symanzik gauge action, and explore four different sets of
lattice spacings, corresponding to $N_t = 4,6,8,10$, and different
spatial sizes, in order to assess the universality class of the
critical point. The continuum extrapolated value of the endpoint
temperature is found to be $T_{\rm RW} = 208(5)$ MeV, i.e. $T_{\rm
  RW}/T_c \sim 1.34(7)$, where $T_c$ is the chiral pseudocritical temperature
at zero chemical potential, while our finite size scaling analysis,
performed on $N_t = 4$ and $N_t = 6$ lattices, provides evidence for a
critical point in the $3d$ Ising universality class.
\end{abstract}

\maketitle

\section{Introduction}
\label{intro}

The remarkable changes expected for the properties of strongly interacting
matter when it is put under extreme conditions are the subject of intense
ongoing theoretical and experimental research. Various parameters of
phenomenological interest enter the description of such extreme conditions,
like temperature, chemical potentials or external background fields. Part of
this research consists in the study of the QCD phase diagram, i.e.~in mapping
the various phases of strongly interacting matter in equilibrium conditions,
and the associated phase transitions and critical points, as a function of
those parameters.

At high temperature confinement and chiral symmetry breaking are expected to
disappear, and QCD is expected to be described in terms of quark and gluon effective degrees
of freedom (Quark-Gluon Plasma). Lattice QCD simulations show that, indeed, a
rapid change of properties takes place around a well defined temperature $T_c$.
There is no compelling reason for expecting a true phase transition, since no
exact symmetry of QCD, which could possibly change its realization at $T_c$, is
known: chiral symmetry is exact only for vanishing quark masses, while the
$Z_3$ center symmetry is exact only in the pure gauge theory, where its
spontaneous breaking is associated to deconfinement.  In fact, lattice
simulations have shown that only a smooth crossover is present in the case of
physical quark masses, at a temperature $T_c \sim 155$
MeV~\cite{aefks,afks,betal,tchot,tchot2}.

The situation could be different in the presence of other external parameters.
In particular, the crossover could turn into a real transition for large enough
baryon chemical potential $\mu_B$, starting from a critical endpoint in the
$T-\mu_B$ plane. Such a critical point, and the associated critical behavior
around it, could have a huge impact on strong interactions phenomenology, so
that large theoretical and experimental efforts are being dedicated to prove
its existence and locate it. Unfortunately, numerical progress by lattice QCD
simulations is strongly hindered by the sign problem affecting the path
integral formulation at non-zero baryon chemical potential.

There are, however, well defined locations, in an extended QCD phase diagram,
where exact symmetries are known for any value of the quark masses.  Critical
points associated with their spontaneous symmetry breaking are predicted to
exist and can be investigated by standard lattice simulations.  This is the
case of QCD with a purely imaginary baryon chemical
potential~\cite{alford,lomb99,fp1,dl1}, the partition function of which is
\begin{equation}
Z(T,\theta_B) = {\rm Tr} \left( e^{-\frac{H}{T}}\ e^{i \theta_B B} \right)
\end{equation}
where $H$ is the QCD Hamiltonian, $B$ is the baryon charge and $\theta_B = {\rm
Im}(\mu_B)/T$. All physical states of the theory, over which the trace is
taken, are globally color neutral and carry an integer valued baryon charge
$B$, hence $Z$ is $2 \pi$-periodic in $\theta_B$, or alternatively $2
\pi/N_c$-periodic in $\theta_q = {\rm Im}(\mu_q)/T$, where $\mu_q = \mu_B /
N_c$ is the quark chemical potential and $N_c$ is the number of colors. That
can also be proven by making use of center transformations in the path-integral
formulation of the partition function, as we review in Section~\ref{sec2}.

On the other hand, in the high-$T$ phase, quarks, which carry a baryon charge
$1/N_c$, become the effective degrees of freedom propagating through the
thermal medium: modes which are $2 \pi$-periodic in $\theta_q$, hence $2\pi
N_c$ periodic in $\theta_B$, appear in the functional dependence of the
partition function. As a consequence, the $2 \pi$ periodicity in $\theta_B$ is
possible only through the appearance of a non-analytic behavior in
$Z(T,\theta_B)$, associated with first order phase transition lines present for
$\theta_B = \pi$ or odd multiples of it, which are known as Roberge-Weiss
(RW) transitions~\cite{rw} and have been widely studied by lattice QCD
simulations~\cite{fp1, dl1, ddl07, cea2009, FMRW, CGFMRW, OPRW, cea2012,
PP_wilson, alexandru, wumeng, wumeng2, nf2BFEPS, nagata15, makiyama16, cuteri, nf2PP}.

In correspondence with such points, analogously to what happens when $\theta_B$
is a multiple of $2 \pi$, the theory is invariant under charge conjugation, but
contrary to that case charge conjugation is spontaneously broken at high $T$,
where the system develops a non-zero expectation value for the imaginary part
of the baryon number density: the temperature $T_{\rm RW}$ where the
spontaneous breaking takes place is precisely the endpoint of the Roberge-Weiss
first order transition lines. An alternative point of view about the same
transition is to look at it as a quantum (i.e. zero temperature) transition,
with an associated spontaneous breaking of charge conjugation, driven by the
compactification of one of the spatial directions beyond a critical
size $L_C = 1/T_{\rm RW}$ (finite size transition
\cite{finitesize,LPP}). Since charge conjugation is a $Z_2$ symmetry, one
expects a $3d$-Ising universality class if the transition is second order, or
alternatively a first order transition with the development of a latent heat.

The temperature $T_{\rm RW}$ and the critical behavior to which it is related
represent universal properties of strong interactions, directly related to the
change in the effective degrees of freedom propagating in the thermal medium,
hence to deconfinement. They can be carefully studied by lattice QCD
simulations, since the path integral measure is real and positive for imaginary
chemical potentials. Despite being related to a critical point located in an
unphysical region of the QCD phase diagram, their importance and relevance to a
full understanding of strong interactions stems from various considerations:

\noindent
{\em i)} The RW endpoint may influence physics in a critical region around it.
Moreover, if at the RW endpoint a first order transition is present, the
endpoint is actually a triple point, with further departing first order lines,
the endpoints of which may be even closer to the $\mu_B = 0$ axis, with more
interesting consequences.

\noindent
{\em ii)} Early studies have shown that the RW endpoint transition is first
order for small quark masses, second order for intermediate masses, and again
first order for large masses; the three regions are separated by two
tricritical points~\cite{FMRW, CGFMRW, OPRW}. The emergence of this interesting
structure has induced many further studies in effective models \cite{model-rw,
Sakai:2009dv, sakai2, sakai3, sakai4, holorw, holorw2, morita, weise, pagura,
buballa, kp13, rw-2color} which try to reproduce the essential features of QCD.
Moreover, interesting proposals have been made on the connection of this phase
structure with that present at $\mu_B$ = 0 (the so-called Columbia plot) and on
the possibility to exploit the whole phase structure at imaginary chemical
potential in order to clarify currently open issues on the phase structure at
$\mu_B = 0$, like the order of the chiral transition for $N_f =
2$~\cite{nf2BFEPS, nf2PP}.

\noindent
{\em iii)}  Once the RW endpoint has been precisely located, it can be taken
as a test ground to compare the lattice techniques presently used to locate the
critical point at real $\mu_B$, so as to assess their reliability and guide
future research on the subject.

\noindent
{\em iv)} The relation of the RW endpoint to the other symmetries of QCD, which
are present at least in well defined limits of strong interactions, is an
interesting issue by itself, which can help elucidate some fundamental
non-perturbative properties of the theory.

In this paper we study the properties of the RW endpoint by lattice simulations
of QCD with physical quark masses.  Its location $T_{\rm RW}$ is determined for
various lattice spacings, corresponding to temporal extensions $N_t =
4,6,8,10$, and then extrapolated to the continuum limit.  Moreover we are able
to determine its universality class, through a finite size scaling analysis, at
two different lattice spacings, namely $N_t = 4,6$.
Finally, in order to approach the issue of the interconnection
between chiral symmetry and the RW endpoint,
we consider the relation of the 
endpoint location to the analytic continuation of 
the pseudocritical
chiral transition temperature $T_c(\mu_B)$ to imaginary chemical potentials.

The paper is organized as follows. In Section~\ref{sec2} we review the general
framework regarding the RW endpoint in a path-integral approach and present
details about our numerical setup and the observables used to investigate the
critical behavior.  In Section~\ref{sec3} we report on our numerical results
regarding the universality class of the endpoint, the continuum extrapolated
value of $T_{\rm RW}$ and its relation with $T_c(\mu_B)$. Finally, in
Section~\ref{sec4} we draw our conclusions.

\section{General framework and numerical setup}
\label{sec2}

We consider a staggered discretization of the $N_f=2+1$ QCD partition function
in the presence of imaginary quark chemical potentials:
\begin{eqnarray}\label{partfunc}
Z &=& \int \!\mathcal{D}U \,e^{-\mathcal{S}_{Y\!M}} \!\!\!\!\prod_{f=u,\,d,\,s} \!\!\! 
\det{\left({M^{f}_{\textnormal{st}}[U,\mu_{f,I}]}\right)^{1/4}}
\hspace{-0.1cm}, \\
\label{tlsyact}
\mathcal{S}_{Y\!M}&=& - \frac{\beta}{3}\sum_{i, \mu \neq \nu} \left( \frac{5}{6}
W^{1\!\times \! 1}_{i;\,\mu\nu} -
\frac{1}{12} W^{1\!\times \! 2}_{i;\,\mu\nu} \right), \\
\label{fermmatrix}
(M^f_{\textnormal{st}})_{i,\,j}&=&am_f \delta_{i,\,j}+\!\!\sum_{\nu=1}^{4}\frac{\eta_{i;\,\nu}}{2}\nonumber
\left[e^{i a \mu_{f,I}\delta_{\nu,4}}U^{(2)}_{i;\,\nu}\delta_{i,j-\hat{\nu}} \right. \nonumber\\
&-&\left. e^{-i a \mu_{f,I}\delta_{\nu,4}}U^{(2)\dagger}_{i-\hat\nu;\,\nu}\delta_{i,j+\hat\nu}  \right] \, .
\end{eqnarray}
The gauge link variables $U$ are used to construct the tree level improved
Symanzik pure gauge action~\cite{weisz,curci}, $\mathcal{S}_{Y\!M}$, where
$W^{n\!\times \! m}_{i;\,\mu\nu}$ is the trace of the $n\times m$ rectangular
loop constructed along the directions $\mu, \nu$ departing from the $i$ site.
The staggered Dirac operator $(M^f_{\textnormal{st}})_{i,\,j}$, instead, is
built up in terms of the two times stout-smeared~\cite{morning} links
$U^{(2)}_{i;\,\nu}$, in order to reduce taste symmetry violations, with an
isotropic smearing parameter $\rho = 0.15$. As usual, the rooting procedure is
adopted to remove the residual degeneracy of the staggered Dirac operator.

When thermal boundary conditions (periodic/anti-periodic for boson/fermion
fields) are taken in the temporal direction, the temperature of the system is
given by $T = 1/(N_t a)$, where $N_t$ is the number of temporal lattice sites
and $a$ is the lattice spacing, related to the bare parameters of the theory.
For a given number of lattice sites in the temporal direction, we can choose
the simulated temperature by tuning the value of the bare coupling constant
$\beta$ and the quark masses $m_s$ and $m_u = m_d \equiv m_l$, in order to change
the lattice spacing while remaining on a line of constant
physics, where $m_{\pi}\simeq 135\,\mathrm{MeV}$ and $m_s/m_{l}=28.15$; this
line has been determined by a spline interpolation of the results
reported in Refs.~\cite{lattsp1, lattsp2, lattsp3}.

Let us now sketch the structure of the phase diagram at imaginary $\mu_B$. This
has already been done in the introduction, by considering the effective degrees
of freedom at work in the different regimes; now we will proceed through an
analysis of the properties of the path integral.  In the presence of a purely
baryonic chemical potential (i.e. $\mu_Q=0$ and $\mu_S=0$), one has $\mu_u =
\mu_d = \mu_s \equiv \mu_q = \mu_B/3$.  When $\mu_q$ is purely imaginary, its
introduction is equivalent to a global rotation of fermionic boundary
conditions in the temporal direction by an angle $\theta_q = {\rm Im}
(\mu_q)/T$, therefore one expects at least a $2 \pi$-periodicity in $\theta_q$
($2 \pi N_c$ in $\theta_B$). However, the actual periodicity is $2 \pi/N_c$,
since a rotation of the fermionic boundary conditions by that angle is
equivalent to a center transformation on the gauge fields, hence it can be
reabsorbed without modifying the path integral \cite{rw}.

\begin{figure}[htb!]
\includegraphics[width=0.92\columnwidth, clip]{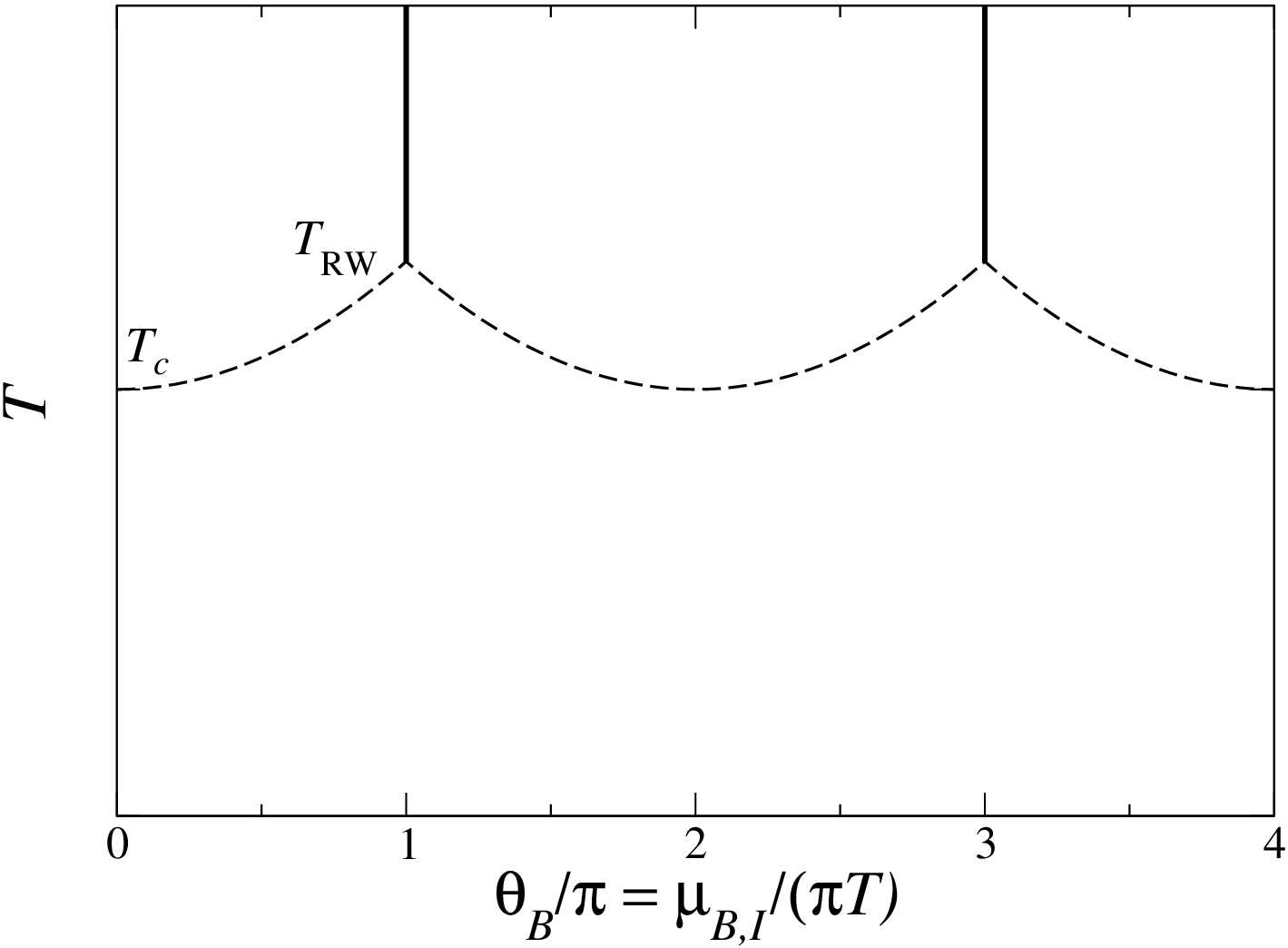}
\caption{Phase diagram of QCD in the presence of an imaginary baryon chemical
potential. The vertical lines represent the Roberge-Weiss transitions taking
place in the high-$T$ regime, while the dashed lines represent the analytic
continuation of the pseudocritical line.}\label{figrw}
\end{figure}

Numerical simulations show that such a periodicity is smoothly realized at low
temperatures \cite{fp1, dl1}. At high $T$, instead, since the Polyakov loop $L$
(trace of the temporal Wilson line normalized by $N_c$) enters the fermionic
determinant expansion multiplied by $\exp (i \theta_q)$, the value of
$\theta_q$ selects the true vacuum among the three different minima of the
Polyakov loop effective potential, which are related to each other by center
transformations. Hence, phase transitions occur as $\theta_q$ crosses the
boundary between two different center sectors, i.e.  for $\theta_q = (2 k +
1)\pi/N_c$ and $k$ integer (in which case $\theta_B$ is an odd multiple of
$\pi$), where $\langle L \rangle$ jumps from one center sector to the
other~\cite{rw}; the phase of $L$ can serve as a possible order parameter in
this case. The $T$-$\theta_q$ phase diagram then consists of a periodic
repetition of first order lines (RW lines) in the high-$T$ regime, which
disappear at low $T$. Therefore they have an endpoint at some temperature
$T_{\rm RW}$, where an exact $Z_2$ symmetry breaks spontaneously. A schematic
view of the diagram is reported in Fig.~\ref{figrw}. 

An alternative order parameter is represented by any of the quark number
densities (where $q=u,d,s$)
\begin{equation}
\langle n_q\rangle \equiv \frac{1}{V_4}\frac{\partial \log Z}{\partial \mu_q}
\end{equation}
where $V_4$ is the four dimensional lattice volume. Since $Z$ is an even
function of $\mu_B$, each $\langle n_q\rangle$ is odd and, for purely imaginary
$\mu_B$, it is purely imaginary as well. Invariance under charge conjugation,
or alternatively oddness and the required $2 \pi$ periodicity in $\theta_B$,
implies that $\langle n_q\rangle$ vanishes for $\theta_B = \pi$ or integer
multiples of it, unless a discontinuity takes place at such points, in
correspondence of a spontaneous breaking of charge conjugation invariance. This
is exactly what happens at the RW lines, so that a non-zero $\langle
n_q\rangle$ signals the onset of the RW transition.

In the following, it will be convenient to consider one particular RW line,
corresponding to $\theta_q = \pi$, for which the imaginary part of the Polyakov
loop, together with the imaginary part of the quark number density, can be
taken as an order parameter. The order parameter susceptibility is then defined
as
\begin{equation}\label{suscdef}
\chi_L \equiv N_tN_s^3\ (\langle ({\rm Im}(L))^2 \rangle - \langle
|{\rm Im}(L)| \rangle^2) \, ,
\end{equation}
where $N_s$ ($N_t$) is the spatial (temporal) size in lattice units. The
susceptibility $\chi_L$ is expected to scale, moving around the endpoint at
fixed $N_t$ and $\theta_q$, as 
\begin{equation}\label{fss}
\chi_L = N_s^{\gamma/\nu}\ \phi (t N_s^{1/\nu}) \, , 
\end{equation}
where $t = (T - T_{\rm RW})/T_{\rm RW}$ is the reduced temperature, which is
proportional to $(\beta - \beta_{\rm RW})$ close enough to the critical point.
That means that the quantity $\chi_L/N_s^{\gamma/\nu}$, measured on different
spatial sizes, should lie on the same curve when plotted against
$(\beta-\beta_{RW})N_s^{1/\nu}$.  Alternatively, we will consider also the
susceptibility of the imaginary part of the quark number density, which is
defined, for every flavor $q$, by
\begin{equation}\label{suscBdef}
\chi_q \equiv N_tN_s^3\ \big(\langle [{\rm Im}(n_q)]^2 \rangle - \langle
  |{\rm Im}(n_q)| \rangle^2\big) \, ,
\end{equation}
and is expected to show a scaling behavior as in Eq.~(\ref{fss}).

\section{Numerical Results}
\label{sec3}

In this Section we present our numerical results, starting from an
analysis of the critical behavior around the RW endpoint transition,
in order to assess its order and universality class on lattices with
$N_t = 4, 6$. Then we will consider also lattices with $N_t = 8, 10$
in order to provide a continuum extrapolated value for $T_{\rm RW}$.

\begin{figure}[t!]
\includegraphics[width=0.92\columnwidth, clip]{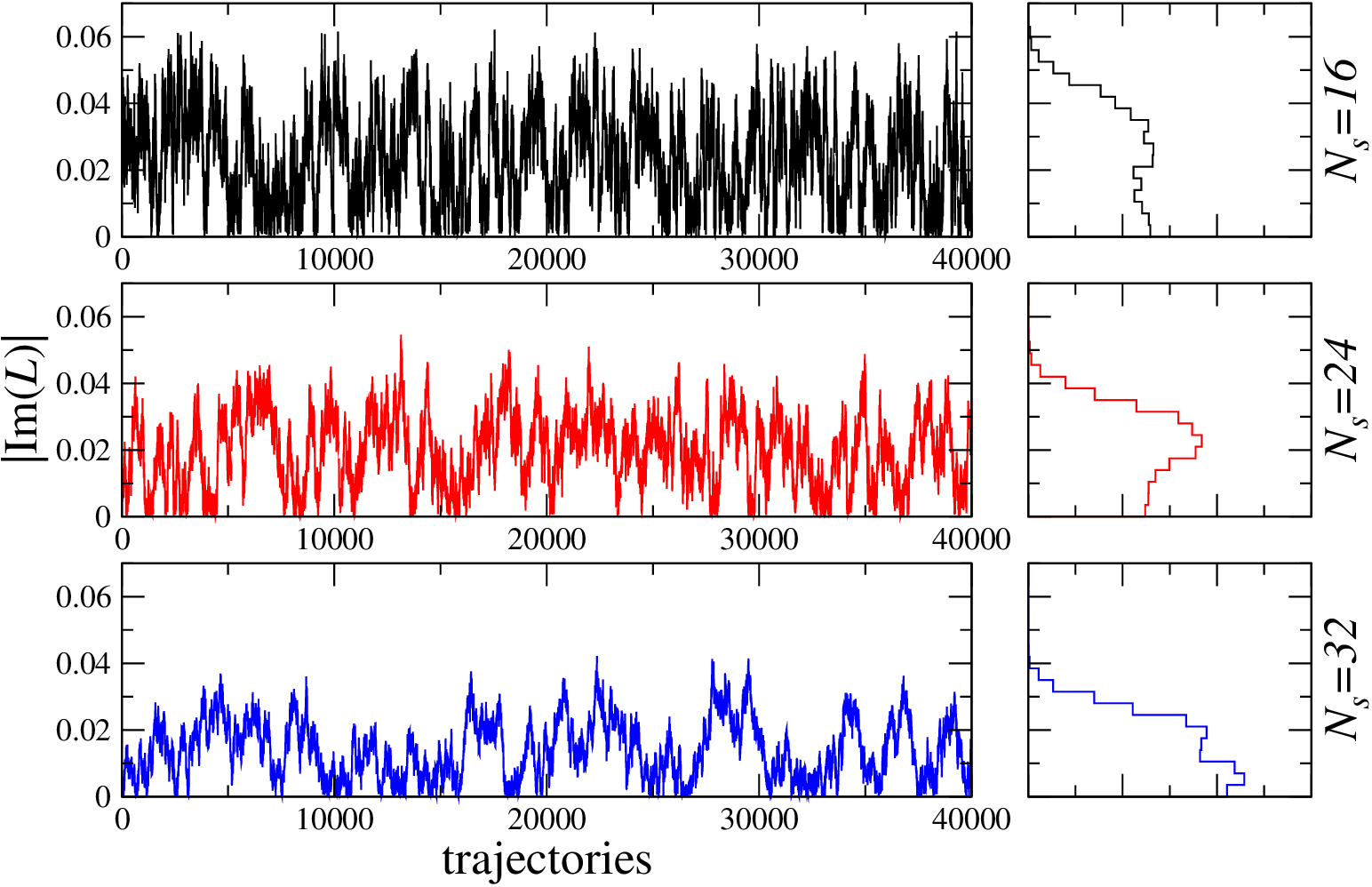}
\caption{Monte Carlo histories of $|{\rm Im}L|$ for $N_t=4$ and 
the $\beta$ values closest to the peak of $\chi_L$, showing the peculiar
features expected near a second order transition: the increase of the
autocorrelation time and the absence of a double peak structure in the
histogram.}\label{fig:histories}
\end{figure}

Since we are interested in studying the behavior near the phase transition,
long time histories are required, to cope with the critical slowing down (see
Fig.~\ref{fig:histories}); for the couplings around the critical value, we used
$\sim 40-50K$ trajectories for each run when performing the finite size
analysis. 

\subsection{Finite size scaling and universality class of the transition}

The effective theory associated with the spontaneous breaking of the charge
conjugation at finite temperature is a three dimensional theory with $Z_2$
symmetry, so the transition can be either first order or second order in the
three-dimensional Ising universality class.  A tricritical scaling is in
principle possible as well; however the tricritical point is just a single
point at the boundary of first and second order regions. As a consequence
(apart from the unlikely case of being exactly on it) tricritical indices
can be observed only as scaling corrections, the ultimate large volume
behavior being either first order or Ising $3d$~\cite{LawSarb, CGFMRW,
potts3d, ouru1}. The critical indices that will be used in the following are
reported for convenience in Table~\ref{tab:critexp}.

\begin{table}[bt!]
\begin{tabular}{|c|c|c|c|c|}
\hline                & $\nu$     & $\gamma$    & $\gamma/\nu$ & $1/\nu$\\
\hline $3D$ Ising     & 0.6301(4) & $1.2372(5)$ & $\sim 1.963$ & $\sim 1.587$ \\
\hline $1^{st}$ Order & 1/3       & 1           & 3            & 3\\
\hline
\end{tabular}
\caption{The critical exponents relevant for this study (see e.g.
\cite{pv_rev, isingcrit}).}\label{tab:critexp}
\end{table}

We will now present the finite size scaling analysis performed to
identify the nature of the transition on lattices with temporal extent
$N_t=4$ and $6$. As previously discussed, we adopt two different
order parameters, namely the imaginary part of the average Polyakov loop and
the quark number density; the former turned out to have smaller
correction to scaling, so we will start our analysis from the study of
the susceptibility $\chi_L$ defined in Eq.~\eqref{suscdef}.

\begin{figure}[ht!]
  \includegraphics[width=0.92\columnwidth, clip]{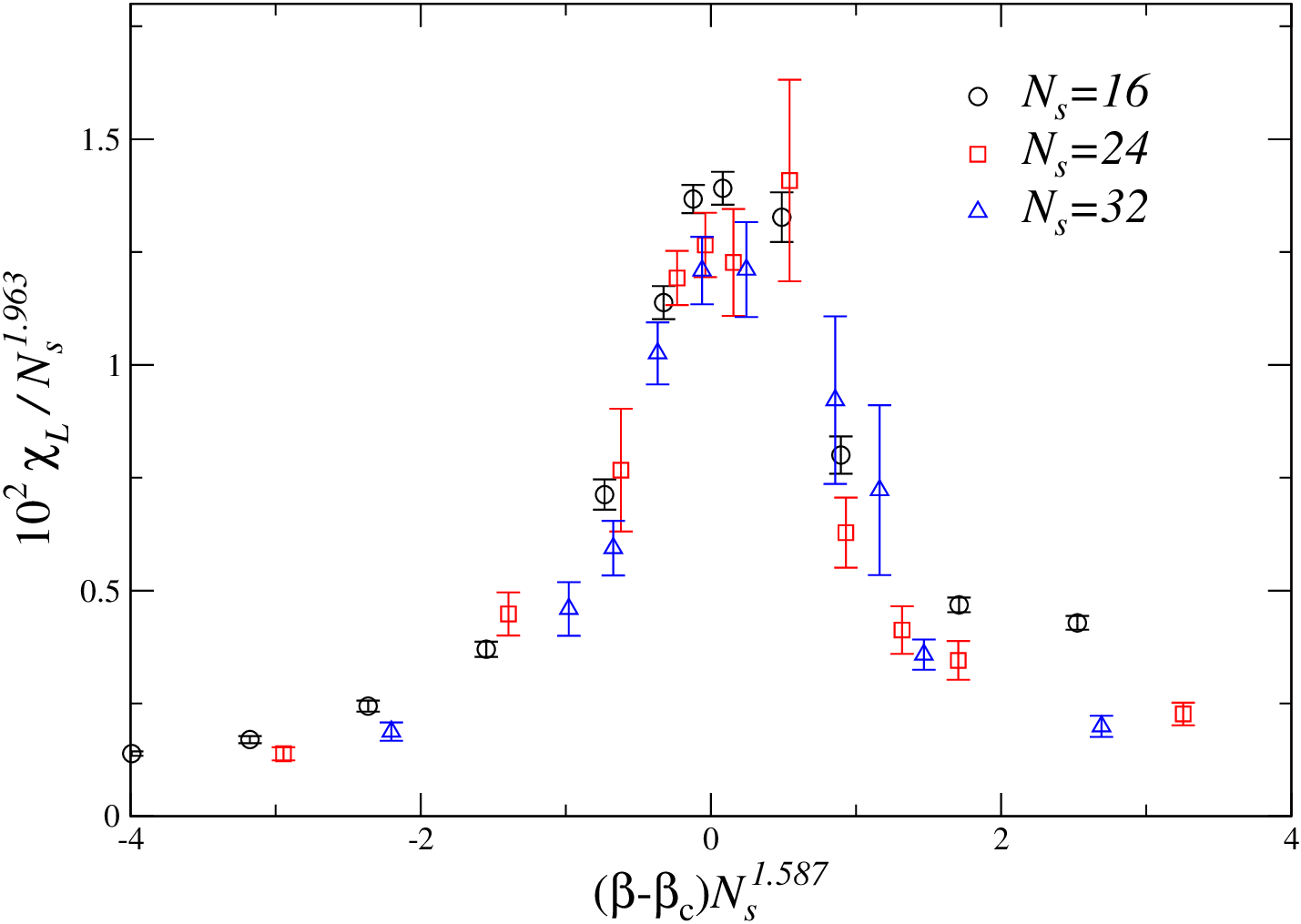}
  \includegraphics[width=0.92\columnwidth, clip]{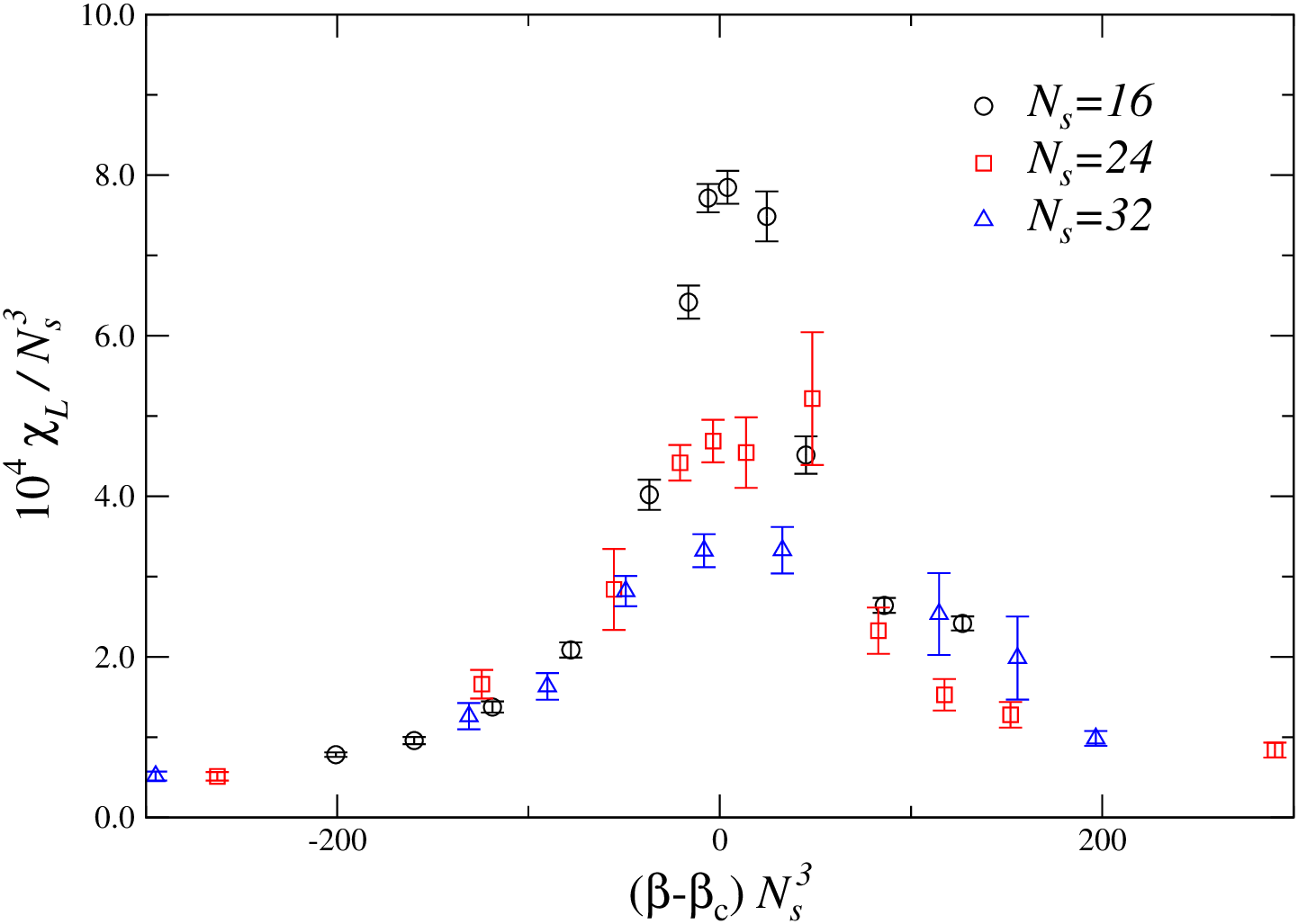}
  \caption{Susceptibility of the imaginary part of the Polyakov loop
    on $N_t=4$ lattices rescaled using the 3d Ising critical indices
    (top) or the first order ones (bottom).}\label{fig:polysusc4}
\end{figure}

\begin{figure}[tb!]
  \includegraphics[width=0.92\columnwidth, clip]{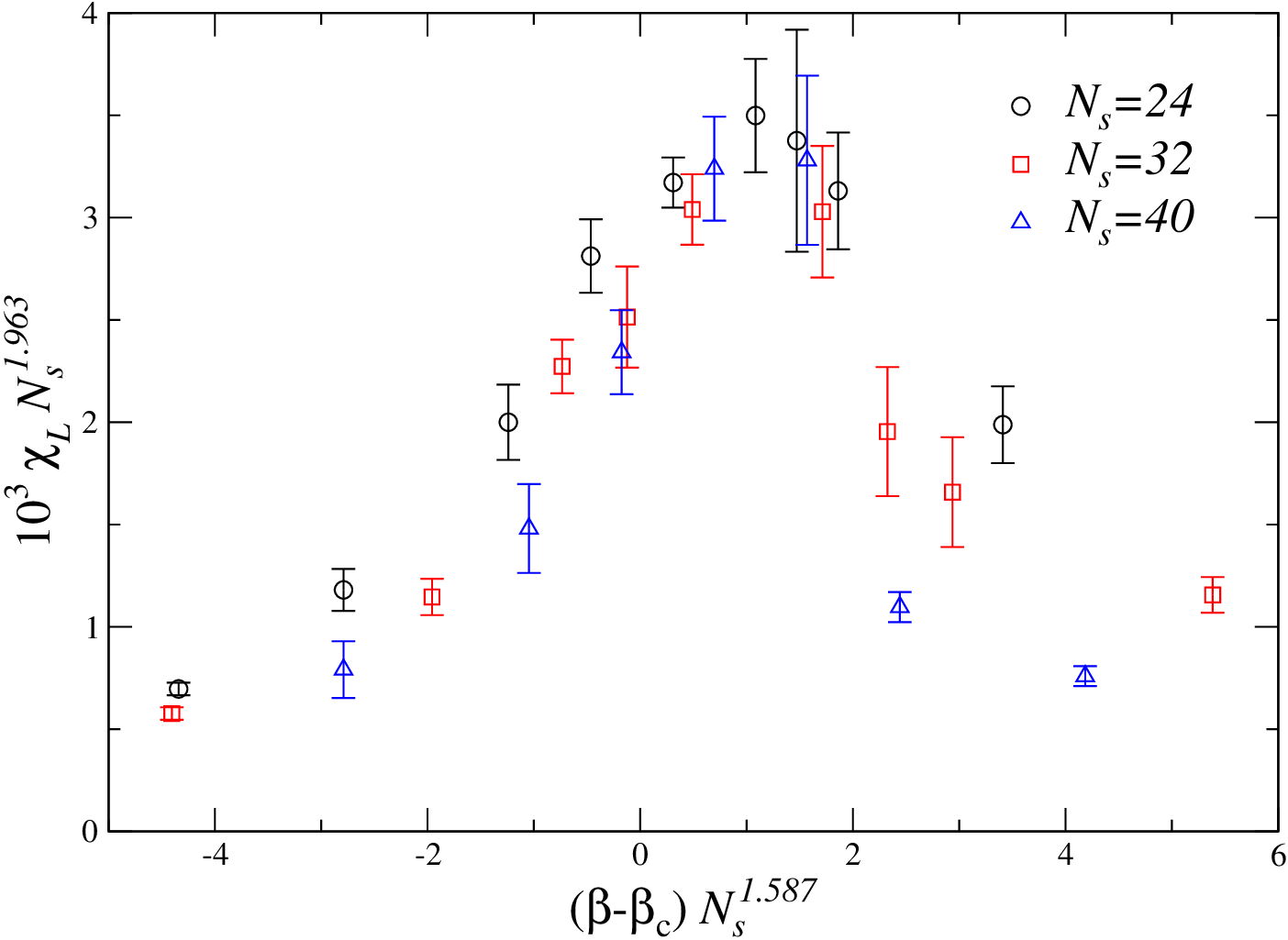}
  \includegraphics[width=0.92\columnwidth, clip]{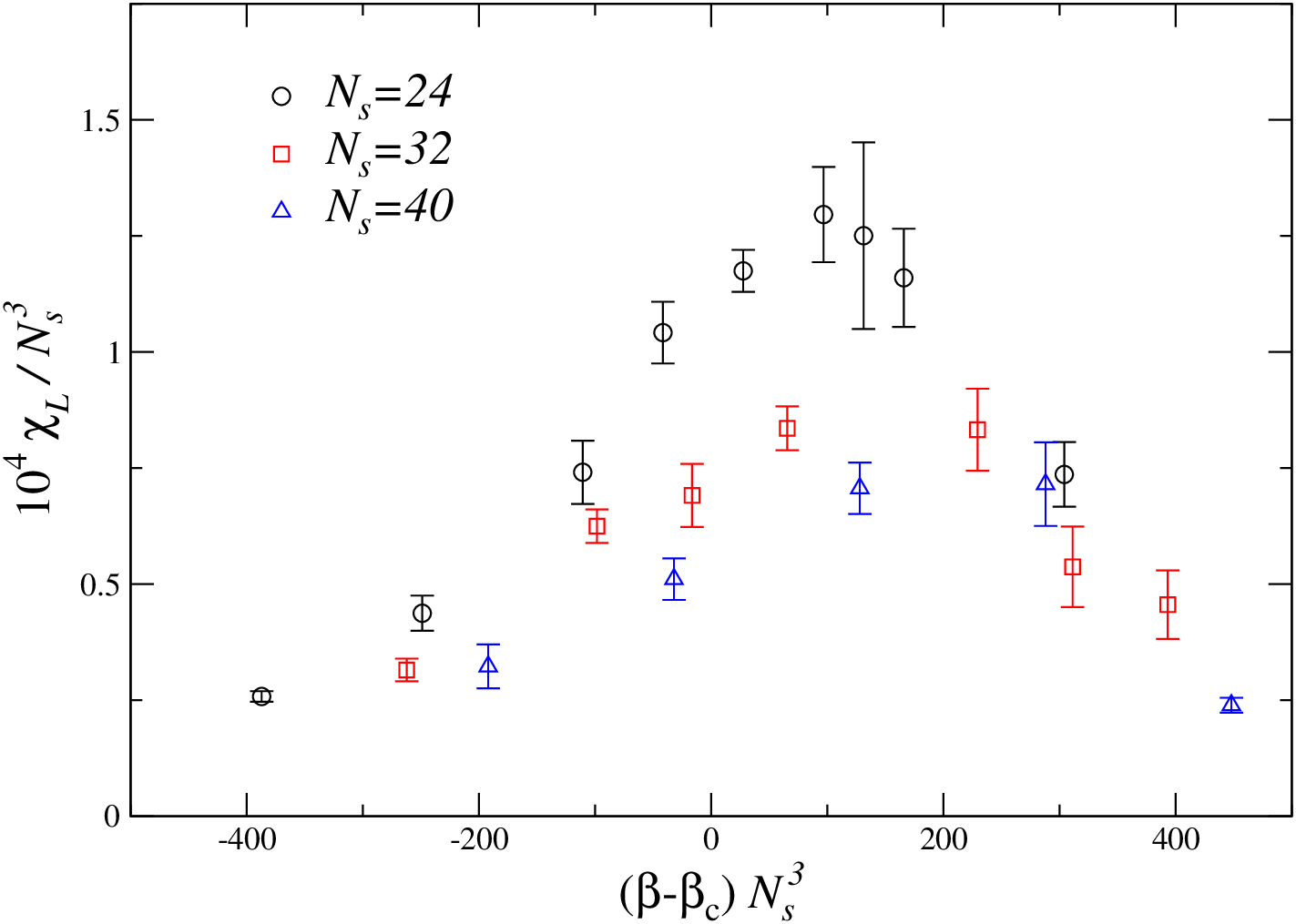}
  \caption{Susceptibility of the imaginary part of the Polyakov loop
    on $N_t=6$ lattices rescaled using the 3d Ising critical indices
    (top) or the first order ones (bottom).}\label{fig:polysusc6}
\end{figure}

Fig.~\ref{fig:polysusc4} shows $\chi_L$ obtained on $N_t=4$ lattices
and rescaled according to Eq.~\eqref{fss}, using alternatively the
critical indices of the $3d$ Ising universality class or those
corresponding to a first order transition (the values used for the
critical coupling are the ones reported in
Table~\ref{tab:critbeta}). Using $3d$ Ising indices the results on
different volumes collapse on top of each other, whereas this is not
the case using first order indices, which strongly indicates that the
transition is second order for $N_t=4$. Note that, since we are
performing simulations on a line of constant physics, the mass
parameters change with $\beta$; it is thus not possible to use
standard reweighting methods \cite{FS1, FS2}. In
Fig.~\ref{fig:polysusc6} we repeat the same analysis using the
Polyakov loop measured on lattices with temporal extent
$N_t=6$. Again, the $3d$-Ising universality class appears to describe
the scaling of the susceptibility of the Polyakov loop significantly
better than a first order, although larger corrections to scaling are
present with respect to the $N_t=4$ case.

A confirmation of the previous analysis comes from the study of the
fourth-order Binder ratio, which in our case is defined as
\begin{equation}
  B_4=\frac{\langle (\mathrm{Im}L)^4\rangle}{\langle
    (\mathrm{Im}L)^2\rangle^2}\ .
\end{equation}
It is easy to show that, in the thermodynamical limit, $B_4\to 3$ in the absence of a
phase transition, while $B_4\to 1$ if a first order transition is present. At 
second order transitions $B_4$ assumes non-trivial values, which are
characteristic of the universal critical behavior associated with the
transition~\cite{Binder1, Binder2, pv_rev}. For the particular case of the
three-dimensional Ising universality class the critical value is
$B_4=1.604(1)$, see Ref.~\cite{isingcrit}.  From these general
properties the following simple procedure follows to locate the critical
endpoint of a line of first order transition: study the behavior of $B_4$ as a
function of the coupling for different values of the lattice size; the endpoint
coupling value will correspond (up to scaling corrections) to the crossing
point of these curves.

\begin{figure}[htb!]
  \includegraphics[width=0.92\columnwidth, clip]{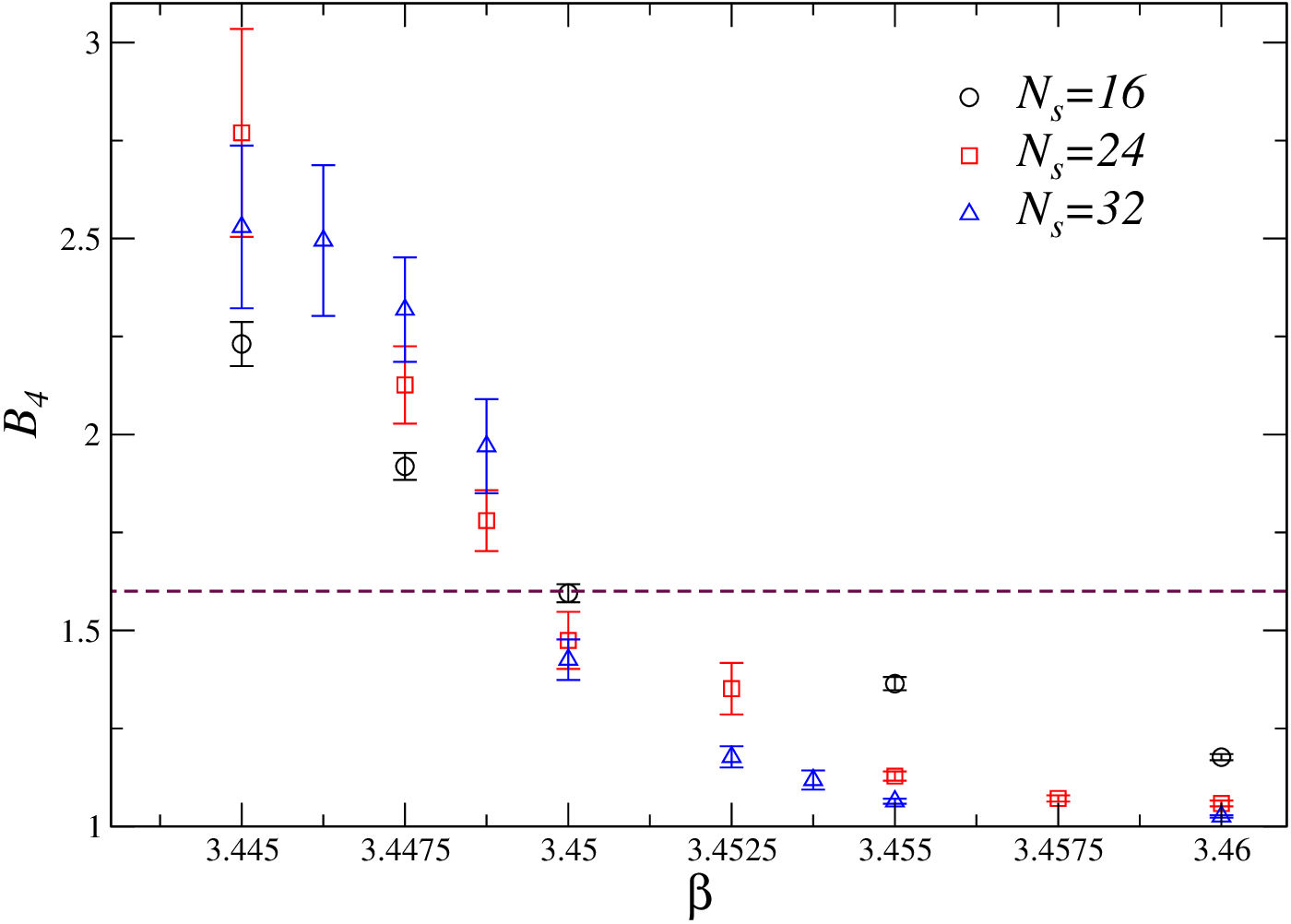}
  \includegraphics[width=0.92\columnwidth, clip]{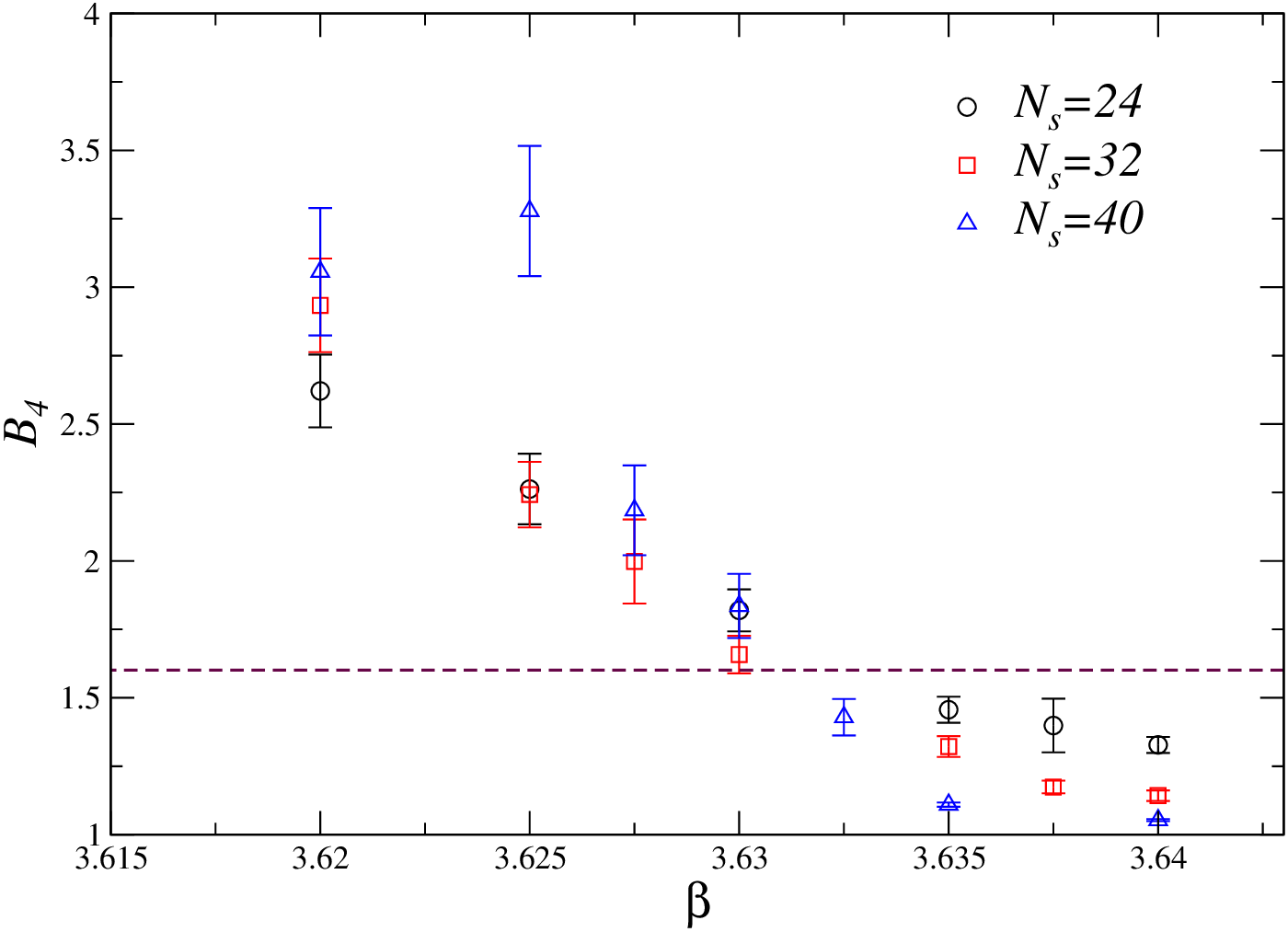}
  \caption{Binder fourth order ratio of the Polyakov loop imaginary
    part computed on $N_t=4$ lattices (top) and $N_t=6$ lattices
    (bottom). The horizontal line denotes the value expected for a
    second order transition of the 3d Ising universality
    class.}\label{fig:polybinder}
\end{figure}

In Fig.~\ref{fig:polybinder} we show the values of $B_4$ in a neighborhood of
the critical coupling at three different volumes both on $N_t=4$ and $N_t = 6$
temporal extent. The behavior of the Binder ratio as a function of $\beta$ is
clearly the one expected at a critical endpoint and the value at the crossing
point is in reasonable agreement with that expected for a transition of the 3d
Ising universality class, while a first order is clearly excluded.

\begin{figure}[htb!]
  \includegraphics[width=0.92\columnwidth, clip]{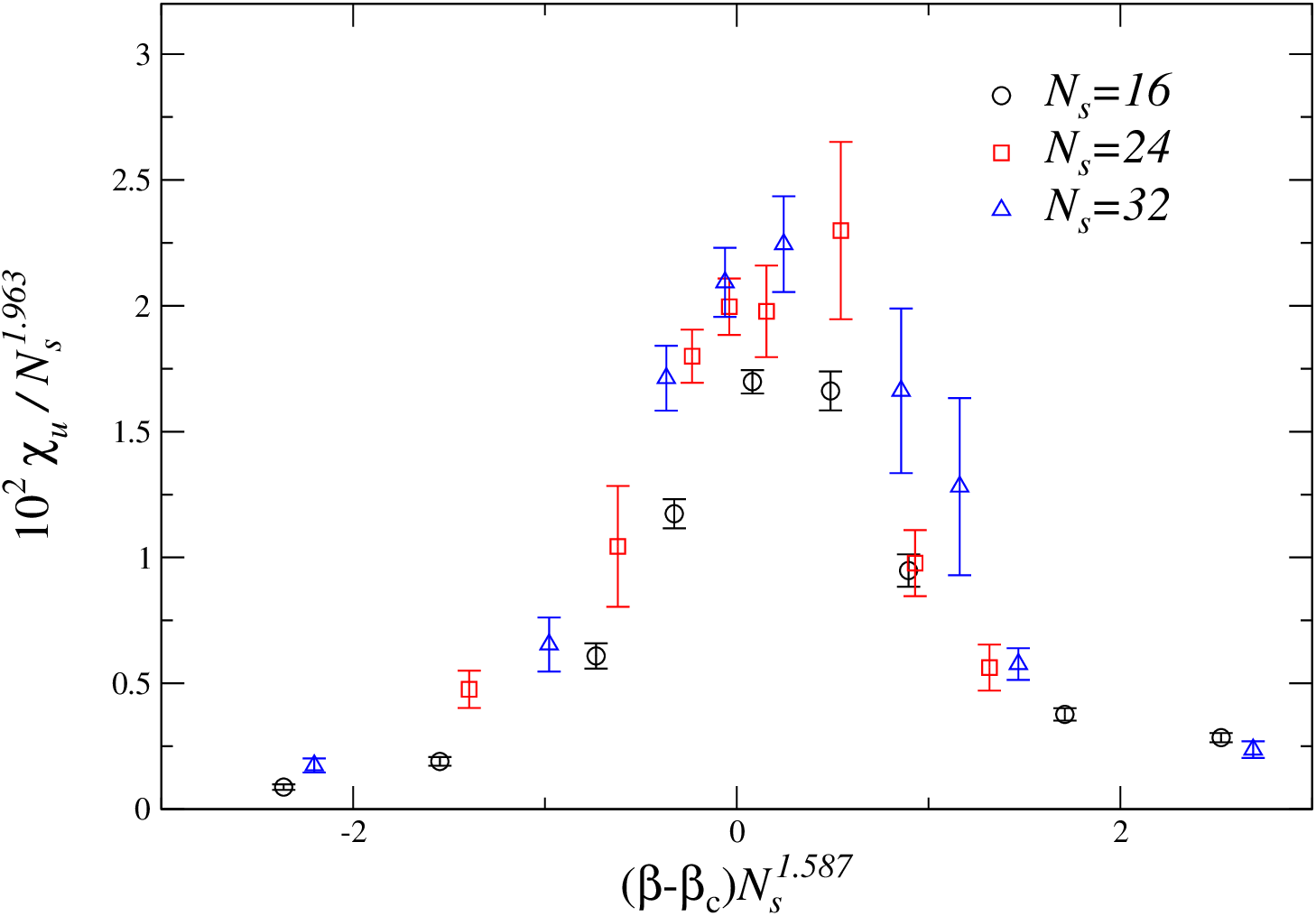}
  \includegraphics[width=0.92\columnwidth, clip]{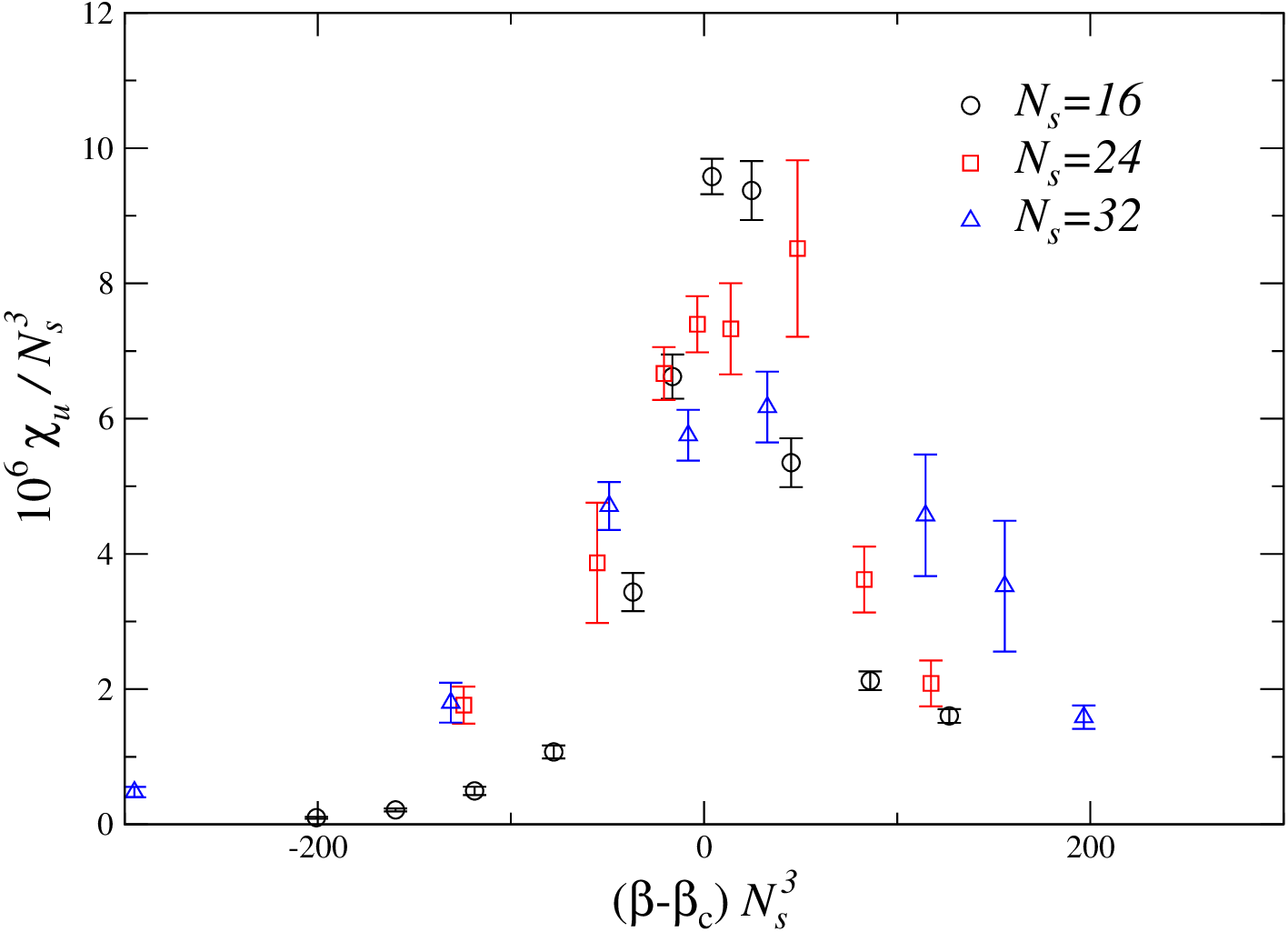}
  \caption{Disconnected susceptibility of the light baryon number
    computed on $N_t=4$ lattices and rescaled with the critical
    exponents of the 3d Ising universality class (top) or
    corresponding to a first order transition
    (bottom).}\label{fig:baryonsusc4}
\end{figure}

The same conclusions are obtained by studying the susceptibility of the $u$
quark number density defined in Eq.~\eqref{suscBdef}, although in this case the
scaling corrections appear to be larger. As an example in
Fig.~\ref{fig:baryonsusc4} we show the behavior of $\chi_u$ on $N_t=4$
lattices, rescaled according to Eq.~\eqref{fss}: again, the 3d-Ising critical
indices are favored.  The case of the strange susceptibility $\chi_s$ is
similar, as well as the $N_t=6$ case.

\subsection{Critical temperature: continuum extrapolated value}

Having established that the RW-transition is second order for
lattices with temporal extent $N_t=4$ and $6$, we now proceed to
estimate the continuum value of $T_{\rm RW}$. To this purpose,
simulations have been performed also on lattices with $N_t=8$ and $10$,
considering a limited number of spatial volumes (one or two) per
simulation setup.

\begin{figure}[htb!]
  \includegraphics[width=0.92\columnwidth, clip]{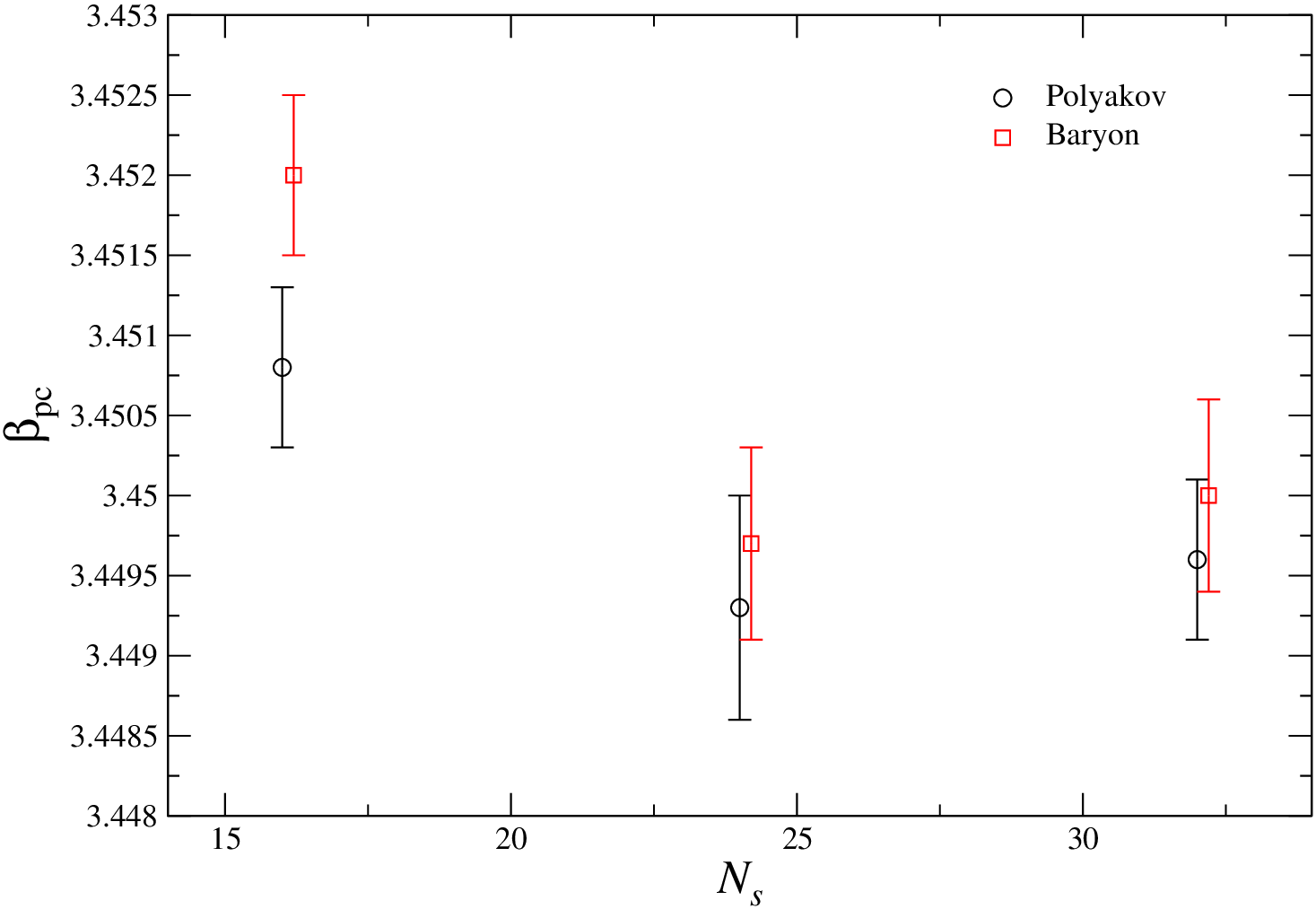}
  \caption{Thermodynamical limit of the pseudocritical coupling
    determined on $N_t=4$ lattices from the maxima of $\chi_L$ and
    $\chi_B$.}\label{fig:critbeta4}
\end{figure}

The pseudocritical value of the coupling has been determined for each
lattice size by estimating the position of the maximum of $\chi_L$ and
$\chi_u$. To this purpose, we have fitted the peak with  a Lorentzian
function:
\begin{equation}
f(\beta)=\frac{a}{1+\left(\beta-\beta_{pc}\right)^2/c^2}\,.
\end{equation}
The results for the large volume limit of $\beta_{pc}$, denoted by
$\beta_c$, are reported in Table~\ref{tab:critbeta}; the error also
takes into account the systematics related to the choice of the fit range. The
volume dependence of the pseudocritical coupling is very mild for lattice with
aspect ratio $4$ or larger, with variations at the level of $0.1\,\%$
in terms of $\beta$ (which become $0.5\,\%$ in terms of temperature), as can be
seen in Fig.~\ref{fig:critbeta4} for the case of the $N_t=4$ lattices. The
pseudocritical couplings determined by using $\chi_L$ or $\chi_u$ have \emph{a
priori} to coincide only in the thermodynamical limit, however in all the cases
the differences between the two determinations are well below $0.1\%$ and,
with the exception of the lattice $4\times 16^3$, they are compatible with each
other at one standard deviation.

\begin{table}
  \begin{tabular}{|l|l|l|l|}
    \hline $N_t$ & $\beta_c$    & $N_s$        &  $a$ (fm) \\
    \hline $4$   & $3.4498(7)$  & $16, 24, 32$ &  $0.2424(6)$ \\
    \hline $6$   & $3.6310(15)$ & $24, 32, 40$ &  $0.1714(3)$ \\
    \hline $8$   & $3.7540(25)$ & $32, 40$     &  $0.1233(3)$ \\
    \hline $10$  & $3.8600(25)$ & $40$         &  $0.0968(2)$ \\
    \hline
\end{tabular}
\caption{Critical values of the coupling for different $N_t$ values (estimated
by using lattices of spatial extent $N_s$) and corresponding values for the
lattice spacing. Only the statistical error of the lattice spacing is reported
in the table, the systematic error is about $2-3\%$ \cite{lattsp1, lattsp2,
lattsp3}.} \label{tab:critbeta}
\end{table}

\begin{figure}[htb!]
  \includegraphics[width=0.92\columnwidth, clip]{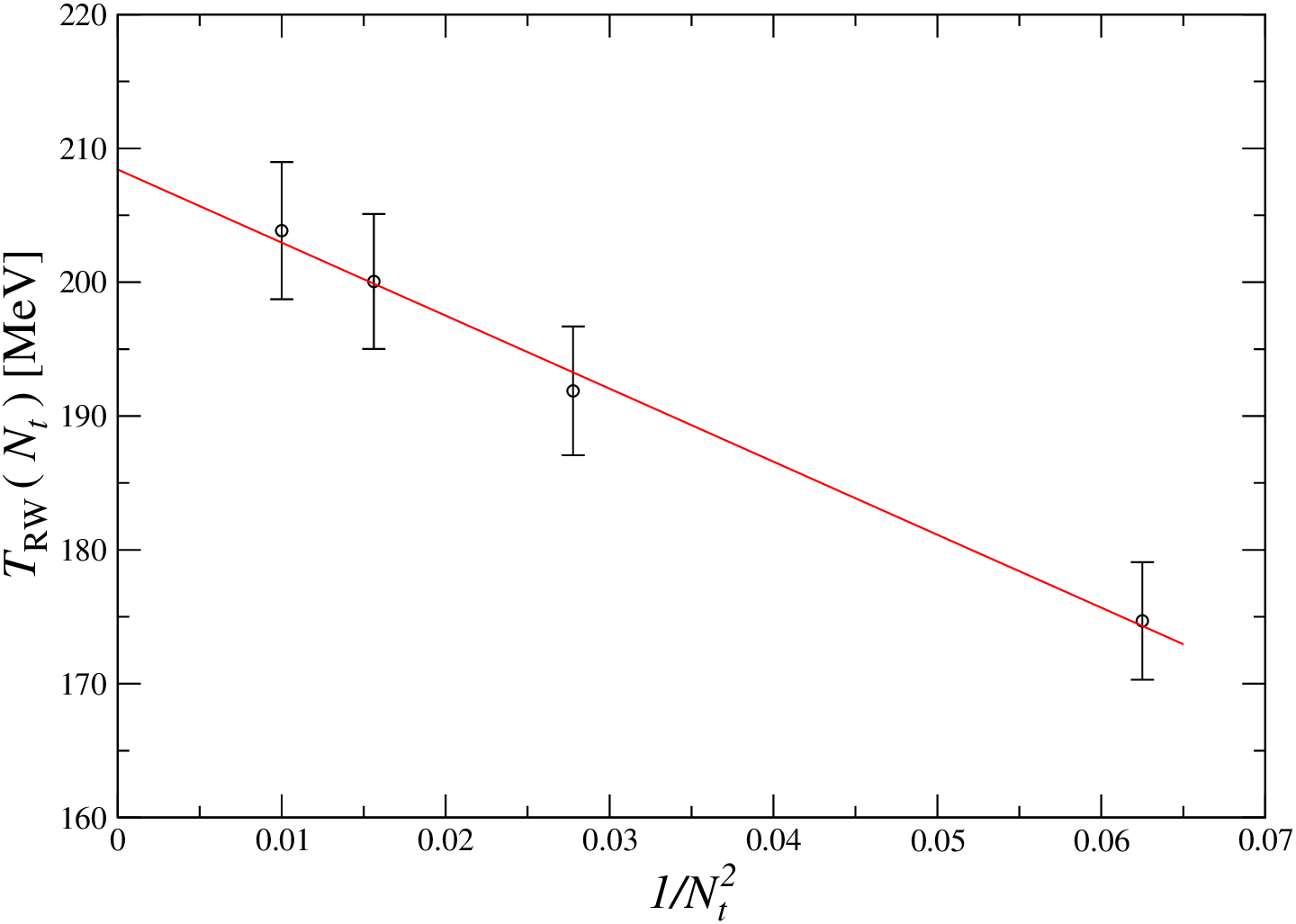}
  \caption{Continuum extrapolation of the critical temperature.}\label{fig:critt}
\end{figure}

In order to convert the critical temperatures to physical units we used the
lattice spacings values reported in Tab.~\ref{tab:critbeta}, which are obtained
by a spline interpolation of the results presented in \cite{lattsp1, lattsp2,
lattsp3}. The systematic uncertainty on these lattice spacings is $2-3\%$
\cite{lattsp1, lattsp2, lattsp3} and this is by far the largest source of error
in the final temperature estimates. The results obtained at the different $N_t$
are plotted in Fig.~\ref{fig:critt} together with the linear fit in $1/N_t^2$,
which describes well the approach to the continuum limit and from which we
extract the value $208(4)\,\mathrm{MeV}$ for the continuum limit of the RW
endpoint temperature. Using as systematical error the difference between this
value and the one obtained using just the three finer lattices, we get
our final estimate $T_{\rm RW}=208(5)\,\mathrm{MeV}$.

\subsection{Relation with the pseudocritical chiral transition line}

An interesting issue that remains to be investigated is the relation
between the RW endpoint and the chiral transition. In particular, the
question can be posed in the following way: does the pseudocritical
line really get to the RW endpoint, as assumed in Fig.~\ref{figrw} and
as suggested by early studies on the subject?

\begin{figure}[htb!]
  \includegraphics[width=0.92\columnwidth, clip]{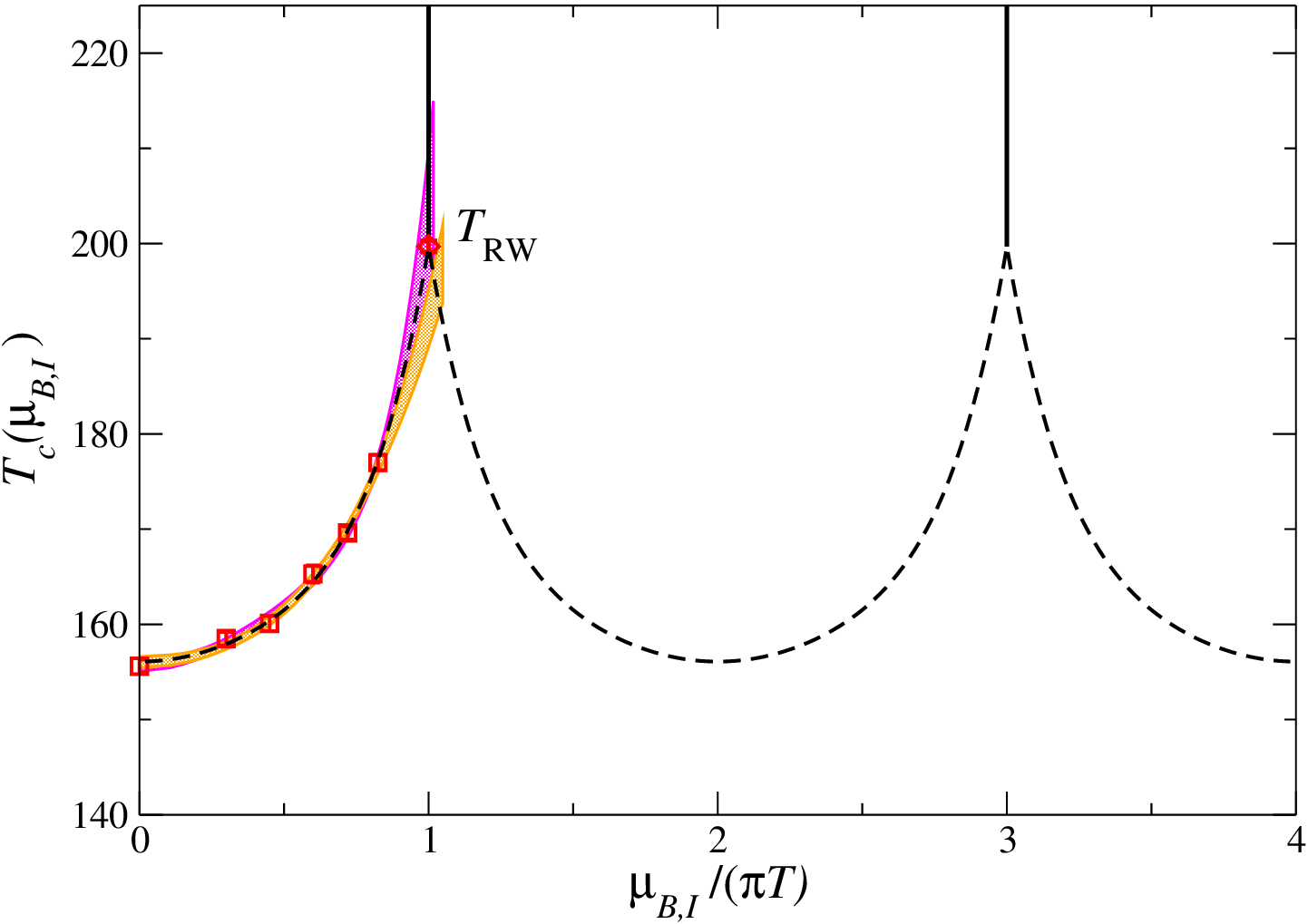}
  \caption{Phase diagram of QCD in the presence of an imaginary
      baryon chemical potential obtained from numerical simulations on
      $N_t=8$ lattices alone. Bands denote fits to polynomials in $\mu_B^2$:
      the orange (longer) band is obtained using terms up to order
      $\mu_B^4$, the violet (shorter) one using up to $\mu_B^6$
      terms.}\label{fig:immu_vero}
\end{figure}

A number of investigations appeared recently, reproducing the
pseudocritical line for imaginary chemical potentials at or close to
the physical point and with the setup of chemical potentials relevant
to the RW endpoint, i.e.~$\mu_s = \mu_l = \mu_B/3$, see
Refs.~\cite{cea_hisq1,corvo,critcurv,cea_hisq2}. A possible way to
approach the question is to try extrapolating the location of the
pseudocritical line up to $\theta_B = \pi$ on the basis of those
determinations. To this aim we considered results for 
$T_c(\theta_B)$ obtained in
Ref.~\cite{critcurv} on $N_t = 8$ lattices and adopting the same
discretization used in the present study. In Fig.~\ref{fig:immu_vero},
we present two different extrapolations of such data, corresponding to
the fit ansatz
\begin{equation}
T_c(\theta_B) = 
T_c (1 + \kappa\, \theta_B^2 + b\, \theta_B^4 
+ c\, \theta_B^6 )
\label{pseudoline}
\end{equation}
with or without the sixth order term included (a simple linear dependence on
$\theta_B^2$ was excluded in Ref.~\cite{critcurv}). In both cases
one gets reasonably close, within errors, to the RW endpoint.

\begin{figure}[htb!]
  \includegraphics[width=0.92\columnwidth, clip]{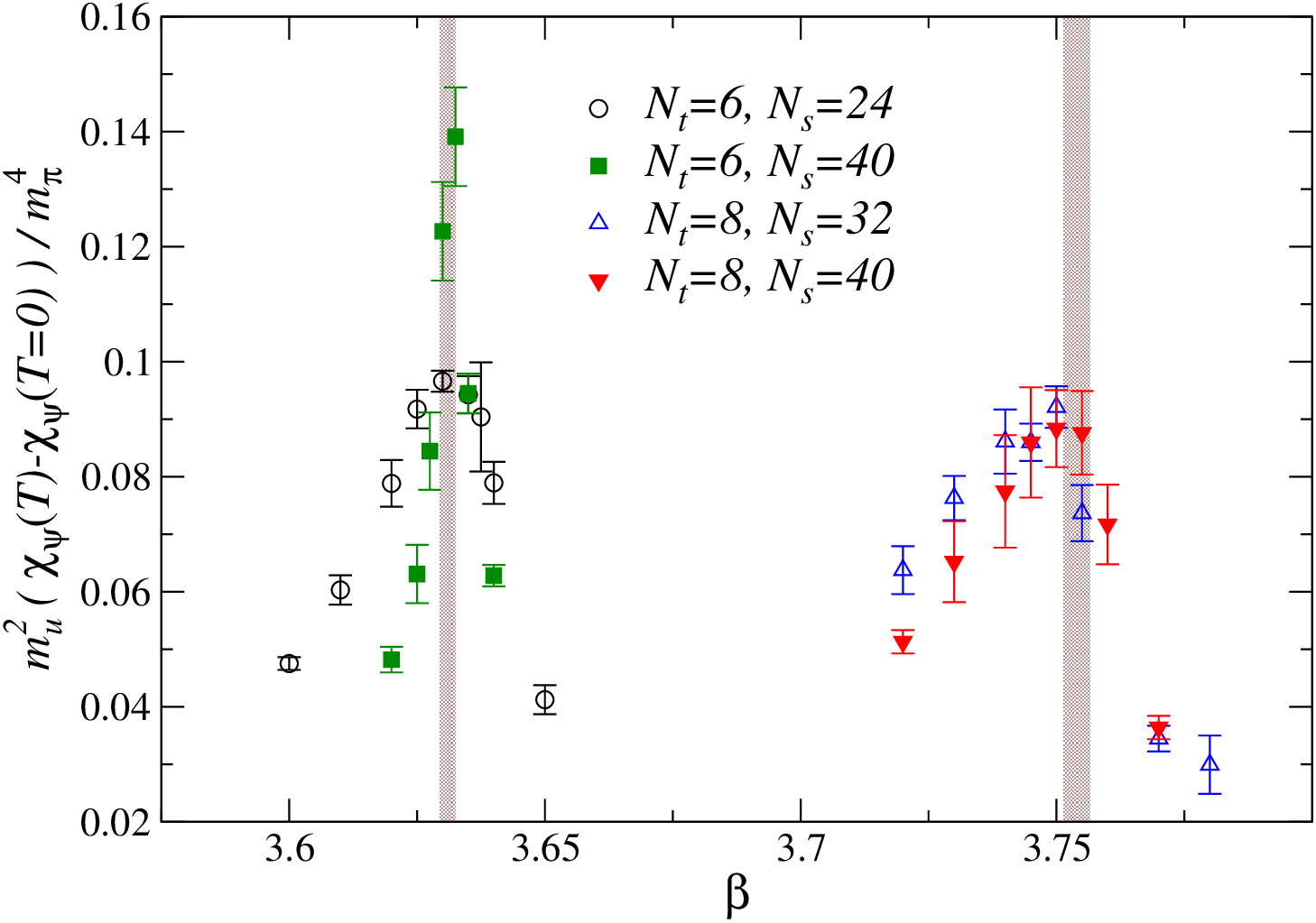}
  \caption{Renormalized light chiral susceptibility on $N_t=6$ and $8$
    lattices.  The vertical bands denote the position of the RW endpoint on
    lattices of the corresponding temporal extent. }\label{fig:chiralsusc}
\end{figure}

Of course, the issue can be checked also directly, by determining the location of
the pseudocritical line exactly at $\theta_B = \pi$. To that aim, in
Fig.~\ref{fig:chiralsusc} we plot the renormalized light chiral
susceptibility (as defined, e.g., in Ref.~\cite{critcurv}) for lattices
with temporal extent $N_t=6, 8$, together with the positions of the RW
endpoint as previously determined on the same lattices. It is clearly
seen that the location of the maxima of the chiral susceptibility
is compatible with the position of the RW endpoints. 
For instance for $N_t = 8$ and $N_s = 32$ we obtain, by
fitting the chiral susceptibility to a Lorentzian peak, $\beta_c =
3.749(3)$, which is at just one standard deviation from the RW
endpoint coupling reported in Table~\ref{tab:critbeta}.

We can thus confirm, within present errors, evidence that the RW endpoint
is located at a point where the analytic continuation of the 
pseudocritical line and the RW first order line meet each other.
To conclude, based on this evidence, 
we have performed a final fit, including terms
up to the sixth order in $\theta_B^2$, which includes the RW endpoint
as a part of the pseudocritical line. The result is the dashed
line reported in Fig.~\ref{fig:immu_vero},
which has been continued also to the other center sectors,
so as to reproduce a realistic version (i.e.~for 
$N_f = 2+1$ QCD with physical quark masses, even if just for $N_t = 8$)
of the phase diagram sketched in Fig.~\ref{figrw}.

\section{Conclusions}
\label{sec4}

We have investigated the properties of the RW endpoint by lattice
simulations of $N_f = 2+1$ QCD with physical quark masses and making
use of two different order parameters for the transition, namely the
imaginary part of the Polyakov line and the imaginary part of the
quark number density, which have led to consistent results.

The temperature of the endpoint, $T_{\rm RW}$, has been determined at four
different values of the lattice temporal extent, $N_t = 4,6,8,10$, from which
we have obtained a continuum extrapolated value $T_{\rm RW} =
208(5)\,\mathrm{MeV}$, where the error includes both statistical and systematic
contributions, stemming mostly from the determination of the physical scale.
That leads to the estimate $T_{\rm RW}/T_c = 1.34(7)$, where the error also
takes into account the systematics involved in the determination of $T_c$,
originating both from the scale setting and from the difficulties in defining a
critical temperature when no real transition is present.  This ratio is
significantly larger than the ones obtained in previous studies; indeed, with
unimproved actions, unphysical quark masses and no extrapolation to the
continuum limit, $T_{\rm RW}$ was typically found to be only about 10\% larger
than $T_c$. The larger value is partially due to the larger 
curvature $\kappa$, and partially 
to the more significant contribution
from non-linear terms in $\mu_B^2$ (see Eq.~\eqref{pseudoline}) which 
are present in the case $\mu_u=\mu_d=\mu_s$ (see Ref.~\cite{critcurv}).

Regarding the order of the transition, our finite size scaling analysis
provides evidence that a second order transition of the $3d$-Ising universality
class takes place, rather than a first order one, at least for $N_t = 4$ and
$N_t = 6$ lattices. Our investigation has been performed at a fixed value of
the pion mass, corresponding to its physical value $m_\pi \simeq 135$ MeV. 

Previous studies on the subject, performed in the $N_f = 2$
theory with both staggered and Wilson fermions, have shown that the
order of the transition changes as a function of $m_\pi$; in
particular, there are two tricritical pion masses, $m_\pi^{\rm
  tric. light}$ and $m_\pi^{\rm tric. heavy}$, and the transition is
second order for $m_\pi^{\rm tric. light} < m_\pi < m_\pi^{\rm
  tric. heavy}$ and first order for lighter or heavier pion masses.
The value of the heavy tricritical mass is typically well above the
GeV scale. The lighter critical pion mass has been found to be
$m_\pi^{\rm tric. light} \sim 400$ MeV for standard staggered fermions
on $N_t = 4$ lattices~\cite{CGFMRW}, 
and around 930 and 680 MeV for standard Wilson
fermions on respectively $N_t = 4$~\cite{PP_wilson} 
and $N_t = 6$~\cite{cuteri} 
lattices.  Given these results, even if we have studied just the physical value
of the pion mass, we can conclude the following: for stout improved staggered
fermions, one has $m_\pi^{\rm tric. light} < 135$ MeV on both the $N_t = 4$ and
$N_t = 6$ lattices. When compared to previous results, that demonstrates the
presence of significant cut-off effects on the values of this tricritical mass,
even when working at fixed $N_t$ but with a different action.  Moreover, based
on the observed tendency of the tricritical mass to decrease with the increase
of $N_t$, we suggest that $m_\pi^{\rm tric. light} $ should be smaller than
$m_{\pi}^{phys}=135$~MeV in the continuum limit, so that the RW endpoint should
be a second order transition in the continuum limit at the physical pion mass.

We must however remark that the mechanism driving the change of nature
of RW endpoint transition, from second to first order as the pion mass
decreases, is still unknown. If such a mechanism is related to the
chiral properties of quarks, unexpected behaviors could occur as the
continuum chiral symmetry group is fully recovered. This is known to
happen, at least for staggered fermions, for lattice spacings well
below those explored in the present study (see Ref.~\cite{axion} for 
a recent investigation about this issue).

Let us spend a few words about what, in our opinion, future
studies should clarify. First of all, one would like to check the
second order nature of the RW endpoint at the physical point on finer
lattices, i.e. for $N_t > 6$. Then, our study with stout improved
staggered fermions should be extended to different values of the pion
mass, in order to locate the values of the tricritical masses
$m_\pi^{\rm tric. light}$ and $m_\pi^{\rm tric. heavy}$ and possibly
extrapolate them to the continuum limit. Such a program, which goes
beyond our present computational capabilities, would clarify the
universal properties of the only critical point of strong interactions
(in the presence of finite quark masses) that one can predict \emph{a priori},
based on the known symmetries of QCD.

Finally, another open issue regards the relation of the RW critical point
to those predicted in well defined limits of QCD.
The relation to the deconfinement transition present in the quenched
case is obvious, since the two transitions trivially coincide in this
case and are both related to center symmetry. The relation to the
chiral transition in the limit of massless quarks is far less
trivial. Suppose to move (varying the temperature) along the line
$\theta_B = \pi$ in the presence of massless quarks; in principle one
expects two different critical temperatures, one at which chiral
symmetry is restored, $T_\chi$, and one at which the $Z_2$ charge
conjugation symmetry spontaneously breaks, $T_{\rm RW}$. What is the
relation between $T_\chi$ and $T_{\rm RW}$? Our present results at finite
quark masses prove that the location of the 
peak of the renormalized chiral
susceptibility coincides, within errors, with $T_{\rm RW}$,
see Fig.~\ref{fig:chiralsusc}, so that the analytic continuation
of the pseudocritical line meets the RW line at its endpoint.
However, in order to obtain
a definite answer, the issue should be 
explored while approaching the chiral
limit: this is something which goes beyond the purpose of the present study
and is left to future investigations.

\acknowledgments

It is a pleasure to thank P.~de Forcrand for very useful comments. 
Numerical simulations have been performed on the BlueGene/Q Fermi
machine at CINECA, based on the Project Iscra-B/CRIBEQCD and on the
agreement between INFN and CINECA (under project INF14\_npqcd),
and on the CSN4 Zefiro cluster of the Scientific Computing
Center at INFN-PISA. FN acknowledges financial support from the INFN SUMA 
project.  FS
received funding from the European Research Council under the European
Community Seventh Framework Programme (FP7/2007-2013) ERC grant
agreement No 279757.

\end{document}